\documentclass[prd,aps,twocolumn,nofootinbib,floats,letterpaper,floatfix,groupedaddress,eqsecnum]{revtex4}
\usepackage{epsfig}
\usepackage{amssymb,amsmath,dcolumn}

\begin{document}

\title{Ergoregion instability of black hole mimickers\footnote{Black Holes in General Relativity and String Theory\\
		 August 24-30 2008, Veli Lo\v{s}inj, Croatia\\}}

\author{Paolo Pani} \email{paolo.pani@ca.infn.it}
\affiliation{Dipartimento di Fisica, Universit\`a di Cagliari, and INFN sezione di Cagliari, Cittadella Universitaria 09042 Monserrato, Italy \\Currently at Centro Multidisciplinar de Astrof\'{\i}sica - CENTRA,
Dept. de F\'{\i}sica, Instituto Superior T\'ecnico, Av. Rovisco
Pais 1, 1049-001 Lisboa, Portugal }

\author{Vitor Cardoso} \email{vcardoso@fisica.ist.utl.pt}
\affiliation{Centro Multidisciplinar de Astrof\'{\i}sica - CENTRA,
Dept. de F\'{\i}sica, Instituto Superior T\'ecnico, Av. Rovisco
Pais 1, 1049-001 Lisboa, Portugal \& \\  Department of
Physics and Astronomy, The University of Mississippi, University,
MS 38677-1848, USA}

\author{Mariano Cadoni} \email{mariano.cadoni@ca.infn.it}
\affiliation{Dipartimento di Fisica, Universit\`a di Cagliari, and INFN sezione di Cagliari, Cittadella Universitaria 09042 Monserrato, Italy}

\author{Marco Cavagli\`a} \email{cavaglia@phy.olemiss.edu}
\affiliation{Department of Physics and Astronomy, The University of Mississippi, University, MS 38677-1848,
USA}

%%%%%%%%%%%%%%%%%%%%%%%%%%%%%%%%%%%%%%%%%%%%%%%%
%%%%%%%%%%%%%%%%%%%%%%%%%%%%%%%%%%%%%%%%%%%%%%%%
\begin{abstract}
Ultra-compact, horizonless objects such as
gravastars, boson stars, wormholes and superspinars can mimick most of the properties of black holes. Here we show that these ``black hole mimickers'' will most likely develop a strong ergoregion instability when rapidly spinning. Instability timescales range between $\sim10^{-5}$s and $\sim$ weeks depending on the object, its mass and its angular momentum. For a wide range of parameters the instability is truly effective. This provides a strong indication that astrophysical ultra-compact objects with large rotation are black holes.
\end{abstract}

\def\t{\times}
\def\be{\begin{equation}}
\def\ee{\end{equation}}
\def\beq{\begin{eqnarray}}
\def\eeq{\end{eqnarray}}

\maketitle
%%%%%%%%%%%%%%%%%%%%%%%%%%%%%%%%%%%%%%%%%%%%%%%%%%%%%%%%%%%%%%%%%%%%%%%%%%%%%%%%%%%%%%%%%%%%%%%%%%%%%%%%%%%%%%%%%
\section{Introduction}
%%%%%%%%%%%%%%%%%%%%%%%%%%%%%%%%%%%%%%%%%%%%%%%%%%%%%%%%%%%%%%%%%%%%%%%%%%%%%%%%%%%%%%%%%%%%%%%%%%%%%%%%%%%%%%%%%
Black holes (BHs) in Einstein-Maxwell theory are characterized by three parameters \cite{Hawking:1971vc}: mass $M$, electric charge $Q$ and angular momentum $J \equiv a M\leqslant M^2$. BHs are thought to be abundant objects in the Universe. Their mass is estimated to vary between $3 M_{\odot}$ and $10^{9.5} M_{\odot}$ or higher \cite{Narayan:2005ie}, their electrical charge is negligible because of the effect of surrounding plasma
\cite{Blandford:1977ds} and their angular momentum is expected to be close to the extremal
limit because of accretion and merger events \cite{Gammie:2003qi}. A non-comprehensive list of some astrophysical BH candidates \cite{Narayan:2005ie,Miller:2004cg,Narayan:2007ks,Wang:2006bz} is shown in Table \ref{tab:DATA}.

Despite the wealth of circumstantial evidence, there is no definite observational proof of the
existence of astrophysical BHs due to the difficulty to detect an event horizon in astrophysical BH candidates \cite{Narayan:2005ie,Abramowicz:2002vt}. Thus astrophysical objects without event horizon, yet observationally indistinguishable from BHs, cannot be excluded a priori. Some of the most viable alternative models describing an ultra-compact astrophysical object include gravastars, boson stars, wormholes and superspinars.\\
Dark energy stars or \emph{gravastars} are compact objects with de Sitter interior and
Schwarzschild exterior \cite{Mazur:2001fv}. These two regions are glued
together by a model-dependent intermediate region. In the original model \cite{Mazur:2001fv} the intermediate region is an ultra-stiff thin shell. Models without shells or discontinuities have also been investigated \cite{Chirenti:2007mk,Cattoen:2005he}.\\
\emph{Boson stars} are macroscopic quantum states which are prevented from undergoing complete
gravitational collapse by Heisenberg uncertainty principle \cite{bosonstars}. Their models differ in the scalar self-interaction potential which also set the allowed maximum compactness for a boson star.\\
An exhaustive description of \emph{wormholes} can be found in the monograph \cite{visserbook} (see also Ref.\ \cite{Lemos:2003jb}). In this work we shall consider particular wormholes which are infinitesimal variations of BH spacetimes. These wormholes may be indistinguishable from ordinary BHs \cite{Damour:2007ap}.\\
\emph{Superspinars} are solutions of the gravitational field equations that violate the Kerr bound.  These geometries could be created by high energy corrections to Einstein gravity such as those present in string-inspired models \cite{Gimon:2007ur}. 

\begin{center}
\begin{table}[ht]
\caption{Mass, $M$, radius, $R$, angular momentum, $J$, and compactness, $\mu=M/R$, for some BH candidates (from \cite{Narayan:2005ie,Miller:2004cg,Narayan:2007ks,Wang:2006bz}). Mass and radius are in solar units.}
\begin{center}
\begin{tabular}{ccccc}
 \hline
 \hline
\textbf{Candidate} & $M$           & $R\times 10^{-5}$  & $J/M^2$  & $\mu=M/R$ \\
 \hline
 GRO J1655-40        & $6.3$                & $1.6 - 2.6$  & $0.65 - 0.80$  &  $0.47 - 0.83$\\
 XTE J1550-564       & $10$                 & $2.1 - 8.4$  & $0.90 - 1.00$  &  $0.25 - 0.99$\\
 GRS 1915+105        & $14$                 & $2.9 - 9.7$  & $0.98 - 1.00$  &  $0.30 - 0.99$\\
 SGR A*              & $4 \times 10^6$      & $\lesssim27$                              & $0.50 - 1.00$  &  $\gtrsim0.31$\\
 \hline
\end{tabular}
\end{center}
\label{tab:DATA} 
\end{table}
\end{center}

The objects described above can be almost as compact as a BH and thus they are virtually indistinguishable from BHs in the Newtonian regime, hence the name ``BH mimickers''. Although exotic these objects provide viable alternatives to astrophysical BHs. BH mimickers being horizonless, no information loss paradox \cite{Hawking:2005kf} arises in these spacetimes. Moreover they can be regular at the origin, avoiding the problem of singularities. By Birkhoff's theorem, the vacuum exterior of a spherically symmetric object is described by the Schwarzschild spacetime. Thus the motion of orbiting objects both around a static BH and around a static ultra-compact object is the same and it makes virtually impossible to discern between a Schwarzschild BH and a static neutral BH mimicker. Instead for rotating objects deviations in the properties of orbiting objects occur. Since BH mimickers are very
compact these deviations occur close to the horizon and are not easily detectable electromagnetically.  To ascertain the
true nature of ultra-compact objects it is thus important to devise observational tests to
distinguish rotating BH mimickers from ordinary Kerr BHs. The traditional way to distinguish a
BH from a neutron star is to measure its mass. If the latter is larger than the Chandrasekhar
limit, the object is believed to be a BH. However, this method cannot be used for the
BH mimickers discussed above, because of their broad mass spectrum. The main difference between a BH and a BH mimicker is the presence of an event horizon in the former. Some indirect experimental methods to detect the event horizon has been proposed \cite{Narayan:1995ic,Broderick:2007ek}. Another very promising observational method to probe the structure of ultra-compact objects is
gravitational wave astronomy. From the gravitational waveform it is expected to detect the presence of an event horizon in the source \cite{Vallisneri:1999nq}. Some other BH mimickers (for example electrically charged quasi-BHs \cite{Lemos:2007yh}) are already ruled out by experiments. Moreover there are evidences that some model for BH mimickers is plagued by a singular behavior in the near-horizon limit \cite{Lemos:2008cv}.

Here, we describe a method originally proposed in \cite{paper1,paper2} for discriminating rotating BH mimickers from ordinary BHs. This method uses the fact that compact rotating objects without event horizon are unstable when an ergoregion is present. This {\it ergoregion instability} appears in any system with ergoregions and no horizons
\cite{friedman}. The origin
of this instability can be traced back to superradiant scattering. In a
scattering process, superradiance occurs when scattered waves have amplitudes larger than
incident waves. This leads to extraction of energy from the scattering body
\cite{zel1,staro1,Bekenstein:1998nt}. Instability may arise whenever this process is allowed to
repeat itself ad infinitum. This happens, for example, when a BH is surrounded by a ``mirror''
that scatters the superradiant wave back to the horizon, amplifying it at each scattering, as in the {\it BH bomb} process \cite{bhbombPress, Cardoso:2004nk}. If
the mirror is inside the ergoregion, superradiance may lead to an inverted BH bomb. Some
superradiant waves escape to infinity carrying positive energy, causing the energy inside the
ergoregion to decrease and eventually generating an instability. This may occur for any
rotating star with an ergoregion: the mirror can be either its surface or, for a star made of
matter non-interacting with the wave, its center. On the other hand BHs could be stable due to the
absorption by the event horizon being larger than superradiant amplification. Indeed Kerr BHs are stable aganist small scalar, electromagnetic and gravitational perturbations \cite{Whiting:1988vc}.

Rapidly rotating stars do possess an ergoregion and thus they are unstable. However typical instability timescales are shown to be larger than the Hubble time \cite{cominsschutz}. Thus the ergoregion instability is too weak to produce any effect on the evolution of stars. This
conclusion changes drastically for BH mimickers due to their compactness \cite{paper1,paper2}. For some of the rotating BH mimickers described above, instability timescales range between $\sim10^{-5}$s and $\sim$ weeks depending on the object, its mass and its angular momentum.

This paper is organized as follows. In Section \ref{sec1} we deal with gravastars and boson stars. We describe rotating models for these objects and discuss their instability timescale. In Section \ref{sec2} a toy model for both rotating wormholes and superspinars is presented. Section \ref{sec:discussion} contains a brief discussion of the results and concludes the paper. Throughout the paper geometrized units ($G=c=1$) are used, except during the discussion of results for rotating boson stars when we set the Newton constant to be $G=0.05/(4\pi)$ as in Ref.\ \cite{Kleihaus:2005me}.
%%%%%%%%%%%%%%%%%%%%%%%%%%%%%%%%%%%%%%%%%%%%%%%%%%%%%%%%%%%%%%%%%%%%%%%%%%%%%%%%%%%%%%%%%%%%%%%%%%%%%%%%%%%%%%%%%
\section{Gravastars and boson stars}\label{sec1}
%%%%%%%%%%%%%%%%%%%%%%%%%%%%%%%%%%%%%%%%%%%%%%%%%%%%%%%%%%%%%%%%%%%%%%%%%%%%%%%%%%%%%%%%%%%%%%%%%%%%%%%%%%%%%%%%%
This section discusses the main properties of gravastars and boson stars as well as the method to compute the ergoregion instability for these objects. For a more detailed discussion see \cite{paper1}.
%%%%%%%%%%%%%%%%%%%%%%%%%%%%%%%%%%%%%%%%%%%%%%%%%%%%%%%%%%%%%%%%%%%%%%%%%%%%%%%
\subsection{\label{subsec:gravastar} Nonrotating Gravastars}
%%%%%%%%%%%%%%%%%%%%%%%%%%%%%%%%%%%%%%%%%%%%%%%%%%%%%%%%%%%%%%%%%%%%%%%%%%%%%%%
Although exact solutions for spinning gravastars are not known, they can be studied in the
limit of slow rotation by perturbing the nonrotating solutions \cite{Hartle:1967he}. This
procedure was used in Ref.~\cite{schutzexistenceergoregion} to study the existence of
ergoregions for ordinary rotating stars with uniform density. In the following, we omit the discussion for the original thin-shell model by Mazur and Mottola \cite{Mazur:2001fv} and we focus on the anisotropic fluid model by Chirenti and Rezzolla \cite{Chirenti:2007mk,Cattoen:2005he}.

\begin{figure}[ht]
\begin{center}
\begin{tabular}{c}
\epsfig{file=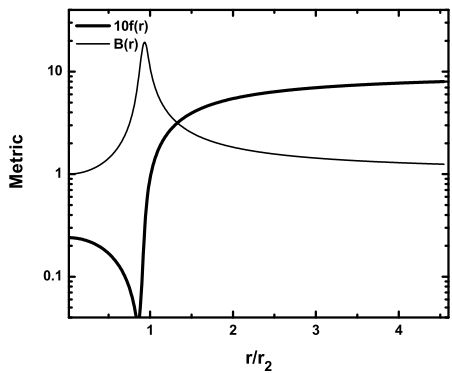,width=200pt}
\end{tabular}
\end{center}
\caption{Metric coefficients for the anisotropic pressure model ($r_2=2.2$, $r_1=1.8$ and $M=1$).}
\label{fig:metricgrava}
\end{figure}

The model assumes a thick shell with continuous profile of anisotropic pressure to avoid the
introduction of an infinitesimally thin shell. The stress-energy tensor is $T^{\mu}{}_{\nu}=
\textrm{diag}[-\rho, p_r, p_t, p_t]$, where $p_r$ and $p_t$ are the radial and tangential
pressures, respectively. The spherical symmetric metric is
\be
ds^2=-f(r)dt^2+B(r)dr^2+r^2d\Omega_2^2 
\label{metricspherical}
\ee
and it consists of three regions: an interior ($r<r_1$) described by a de Sitter metric, an exterior ($r>r_2$) described by the Schwarzschild metric and a model-dependent intermediate ($r_1<r<r_2$) region. In the following we shall indicate with $\delta=r_2-r_1$ the thickness of the intermediate region and with $\mu=M/r_2$ the compactness of the gravastar. In the model by Chirenti and Rezzolla the density function is
\beq
\rho(r) = \left\{ \begin{array}{lll} \rho_0\,, & 0 \le r \le r_1 &\quad \textrm{interior}\nonumber\\
ar^3+br^2+cr+d\,, & r_1 < r < r_2 &\quad \textrm{intermediate}\nonumber\\
0\,, & r_2 \le r                  &\quad \textrm{exterior}
\end{array} \right.\,
\label{rho}
\eeq
where $a$, $b$, $c$ and $d$ are found imposing continuity conditions $\rho(0)=\rho(r_1) = \rho_0$, $\rho(r_2) = \rho'(r_1) = \rho'(r_2) =0$ and $\rho_0$ is found fixing the total mass, M. The metric coefficients are
\be
f=\left(1-\frac{2M}{r_2}\right)
e^{\Gamma(r)-\Gamma(r_2)}\,,\quad  \frac{1}{B}=1-\frac{2m(r)}{r}\,,
\label{lambda}
\ee
where
\be
m(r)=\int_0^r 4\pi r^2\rho d r\,, \hspace{0.5cm} \Gamma(r)= \int_0^r \frac{2m(r)+8\pi r^3 p_r}{r(r-2m(r))}d r\,.
\label{Gamma}
\ee
The above equations and some closure relation, $p_r=p_r(\rho)$, completely determine the structure of the gravastar \cite{Chirenti:2007mk}. The behaviors of the
metric coefficients for a typical gravastar are shown in Fig.~\ref{fig:metricgrava}.
%%%%%%%%%%%%%%%%%%%%%%%%%%%%%%%%%%%%%%%%%%%%%%%%%%%%%%%%%%%%%%%%%%%%%%%%%%%%%%%%%%%%%%%%%%%%%%%%%%%%%%%%%%%%%%%
\subsubsection{\label{sec:ergoregion} Slowly rotating gravastars
and ergoregions}
%%%%%%%%%%%%%%%%%%%%%%%%%%%%%%%%%%%%%%%%%%%%%%%%%%%%%%%%%%%%%%%%%%%%%%%%%%%%%%%%%%%%%%%%%%%%%%%%%%%%%%%%%%%%%%%
Slowly rotating solutions can be
obtained using the method developed in Ref.~\cite{Hartle:1967he}. A rotation of order $\Omega$ gives corrections of order $\Omega^2$ in the diagonal coefficients
of the metric (\ref{metricspherical}) and introduces a non-diagonal term of order $\Omega$, $g_{t\phi}\equiv -\omega g_{\phi\phi}$, where $\phi$ is the azimuthal coordinate and $\omega=\omega(r)$ is the angular velocity of frame dragging. The full metric is
\be
ds^2=-fdt^2+B dr^2+r^2d\theta^2+r^2\sin^2\theta\left(d\phi-\omega dt\right )^2\,,
\label{rotfull}
\ee
where $f$, $B$ and $\omega$ are radial functions.
%
% We consider the anisotropic fluid stress-energy tensor
% %
% \be
% T^{\mu\nu}=(\rho+p_t)U^{\mu}U^{\nu}+p_tg^{\mu\nu}+(p_r-p_t)s^\mu s^\nu \,,
% \label{Tmunu}
% \ee
% %
% where $U^\mu$ and $s^\mu$ are \textbf{XXX} respectively \cite{Chirenti:2007mk}.
% % $U^{\mu}U_{\mu}=-s^{\mu}s_{\mu}=-1$, $U^{\mu}
% % s_{\mu}=U^r=U^{\theta}=0$, $U^{\phi}=\Omega U^t$ and $U^t=\left [-\left (g_{tt}+2\Omega
% % g_{t\phi}+\Omega^2g_{\phi\phi}\right )\right]^{-1/2}$. 
% Equation (\ref{Tmunu}) describes an anisotropic fluid with radial pressure $p_r$ and tangential
% pressure $p_t$, rotating with angular velocity $\Omega$ as measured by an observer at rest in
% the $(t\,,r\,,\theta\,,\phi)$ coordinates. 
If the gravastar rotates rigidly, i.e.\ $\Omega={\rm constant}$, from the $(t,\phi)$ component of Einstein equations we find a differential equation for $\omega(r)$ \cite{paper1}
\be
\omega''+\omega'\left (\frac{4}{r}+\frac{j'}{j}\right ) =16\pi
B(r) (\omega-\Omega) \left (\rho+p_t\right )\,,
\label{zeta}
\ee
where $j\equiv (fB)^{-1/2}$ is evaluated at zeroth order and $\rho$, $p_t$ are given in terms
of the nonrotating geometry. The above equation reduces to the corresponding equation for isotropic fluids \cite{Hartle:1967he}. Solutions of Eq.~(\ref{zeta}) describe rotating gravastars to first order in $\Omega$.

\begin{figure}[ht]
\begin{center}
\begin{tabular}{l}
\epsfig{file=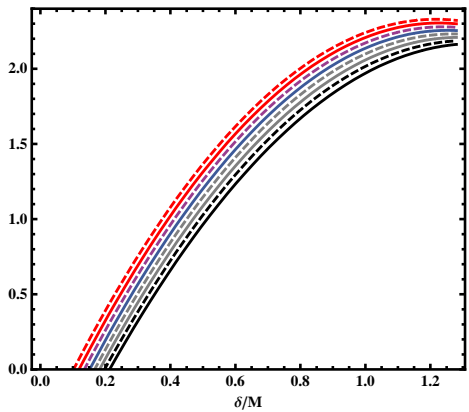,width=180pt,angle=0}
\end{tabular}
\end{center}
\caption{Ergoregion width (in units of M) as function of the thickness, $\delta=r_2-r_1$, for $r_2=2.3$, $M=1$ and for different $J$ values. From top to bottom: $J/M^2=0.95$, $0.90$, $0.85$, $0.80$, $0.75$, $0.70$, $0.65$ and $0.60$. The ergoregion width decreases as $\delta\rightarrow0$.}
\label{fig:delta}
\end{figure}

The ergoregion can be found by computing the surface on which $g_{tt}$ vanishes
\cite{schutzexistenceergoregion}. An approximated relation for the location of the ergoregion in very compact gravastars is
\be
0=-f(r)+\omega^2 r^2\sin^2\theta\,.
\label{eqergo}
\ee
The existence and the boundaries of the ergoregions can be computed from the above equations.
We integrate equation (\ref{zeta}) from the origin with initial conditions $(\Omega-\omega)'=0$
and $(\Omega-\omega)$ finite. The exterior solution satisfies
$\omega=2J/r^3$, where $J$ is the angular momentum of the gravastar. Demanding the
continuity of both $(\Omega-\omega)'$ and $(\Omega-\omega)$, $\Omega$ and $J$ are uniquely
determined. The rotation parameter $\Omega$ depends on the initial condition at the origin.
\begin{figure}[ht]
\begin{center}
\begin{tabular}{l}
\epsfig{file=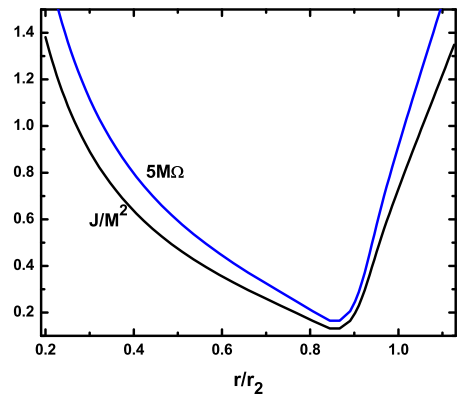,width=180pt,angle=0}
\end{tabular}
\end{center}
\caption{$J/M^2$ and angular frequency
$\Omega$ for the anisotropic pressure model with $r_2=2.2$, $r_1=1.8$ and $M=1$.}
\label{fig:ergoregion}
\end{figure}
Figure \ref{fig:ergoregion} shows the results the gravastar model described in the
previous sections. The ergoregion can be located by drawing an
horizontal line at the desired value of $J/M^2$. The minimum of the curve is the minimum values of $J/M^2$ which are
required for the existence of the ergoregion. Comparison with the results for stars of uniform
density \cite{schutzexistenceergoregion}, shows that ergoregions form more easily around
gravastars due to their higher compactness. The slow-rotation approximation is considered valid for $\Omega/\Omega_K < 1$ where $M\Omega_K=\mu^{3/2}$ is the Keplerian frequency.\\
Depending on the compactness, $\mu$, the angular momentum, $J$, and the thickness, $\delta$, a spinning gravastar does or does not develop an ergoregion. The formation of an ergoregion for rotating gravastar is exhaustively discussed in the whole parameters space in Ref.\ \cite{Chirenti:2008pf}. A delicate issue is the strong dependence on the thickness, $\delta$, which cannot be directly measured by experiments. Figure \ref{fig:delta} shows how the ergoregion width is sensitive to $\delta$.
%%%%%%%%%%%%%%%%%%%%%%%%%%%%%%%%%%%%%%%%%%%%%%%%%%%%%%%%%%%%%%%%%%%%%%%%%%%%%%%%%%%%%%%%%%%%%%%%%%%%%%
\subsection{\label{subsec:bosonstars} Rotating boson stars}
\begin{center}
\begin{figure}[ht]
\begin{tabular}{cc}
\epsfig{file=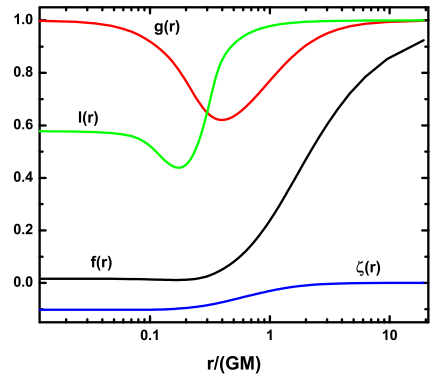,width=190pt,angle=0}\\
\epsfig{file=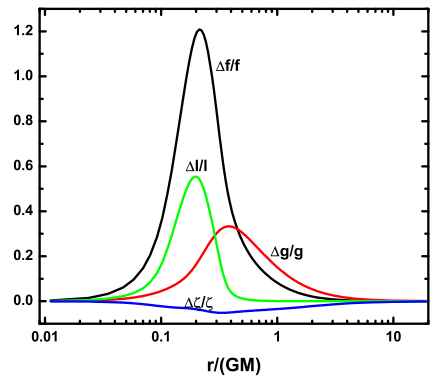,width=190pt,angle=0}
\end{tabular}
\caption{Left panel: Metric coefficients for a rotating boson star along the equatorial
plane, with parameters $n=2$, $b=1.1$, $\lambda=1.0$, $a=2.0$, $J/(GM^2)\sim 0.566$.  Right
panel: Fractional difference of the metric potentials between $\theta=\pi/2$ and
$\theta=\pi/4$ for the same star.}
\label{fig:metricrotboson}
\end{figure}
\end{center}

%%%%%%%%%%%%%%%%%%%%%%%%%%%%%%%%%%%%%%%%%%%%%%%%%%%%%%%%%%%%%%%%%%%%%%%%%%%%%%%%%%%%%%%%%%%%%%%%%%%%%%
A example of rotating boson star is the model by Kleihaus, Kunz, List and
Schaffer (KKLS) \cite{Kleihaus:2005me}. The KKLS solution is
based on the Lagrangian for a self-interacting complex scalar field
\be {\cal L}_{KKLS}=-\frac{1}{2} g^{\mu\nu}\left( \Phi_{, \, \mu}^*
\Phi_{, \, \nu} + \Phi _ {, \, \nu}^* \Phi _{, \, \mu} \right) - U(
\left| \Phi \right|)\,, \label{KKL} \ee
where $U(|\Phi|)=\lambda |\Phi|^2(|\Phi|^4-a|\Phi|^2+b)$. The mass of the boson is given by
$m_{\rm B}=\sqrt{\lambda b}$. The ansatz for the axisymmetric spacetime is
\be
ds^2 =- f dt^2 + \frac{kg}{f} \, \biggl[dr^2 + r^2 \, d\theta^2+\frac{r^2 \, \sin^2 \theta}{g} \,\left( d \varphi-\zeta(r) \, dt \right)^2 \biggr]\label{ansatzg}
\ee
and $\Phi=\phi~e^{ i\omega_s t +i n \varphi}$, where the metric components and the real
function $\phi$ depend only on $r$ and $\theta$. The requirement that $\Phi$ is single-valued
implies $n=0,\pm 1,\pm 2,\dots$. The solution has spherical symmetry for $n=0$ and axial
symmetry otherwise. Since the Lagrangian density is invariant under a global $U(1)$ transformation,
the current, $j^{\mu}=-i\Phi^*\partial^{\mu}\Phi+{\rm c.c.}$, is conserved and it is associated to a charge $Q$, satisfying the quantization condition with the angular momentum $J=nQ$ \cite{schunck}. The numerical procedure to extract the metric and the scalar field is described in Ref.\ \cite{Kleihaus:2005me}. Throughout the paper we will consider solutions with $n=2$, $b=1.1$, $\lambda=1.0$, $a=2.0$ and different values of $(J\,,M)$ corresponding to $J/(GM^2)\sim 0.566$, $0.731$ and $0.858$. In Fig.~\ref{fig:metricrotboson} the metric functions for a boson star along the equatorial plane are shown. By computing the coefficient $g_{tt}$ one can prove that boson stars
develop ergoregions deeply inside the star. For this particular choice of parameters, the
ergoregion extends from $r/(GM)\sim 0.0471$ to $0.770$. A more complete discussion on the
ergoregions of rotating boson stars can be found in Ref.~\cite{Kleihaus:2005me}.
%
% \begin{figure}
% \begin{center}
% \begin{tabular}{l}
% \epsfig{file=RotatingBosongtt.eps, width=210pt,angle=0} 
% \end{tabular}
% \end{center}
% \caption{The $g_{tt}$ metric coefficient for a boson star
% with  $J/(GM^2)\sim 0.566$ at its equator. The ergoregion is
% identified by the region inside the dotted vertical lines and
% extends from $r/(GM)\sim 0.047$ to $0.770$.}
% \label{fig:metricrotbosongtt}
% \end{figure}
% %
%%%%%%%%%%%%%%%%%%%%%%%%%%%%%%%%%%%%%%%%%%%%%%%%%%%%%%%%%%%%%%%%%%%%%%%%%%%%%%%%%%%%%%%%%%%%%%%%%%%%
\subsection{Ergoregion instability for rotating gravastars and boson stars}
\label{sec:ergoinst}
%%%%%%%%%%%%%%%%%%%%%%%%%%%%%%%%%%%%%%%%%%%%%%%%%%%%%%%%%%%%%%%%%%%%%%%%%%%%%%%%%%%%%%%%%%%%%%%%%%%%
The stability of gravastars and boson stars can be studied perturbatively by considering small
deviations around equilibrium. Due to the difficulty of handling gravitational perturbations for rotating objects, the
calculations below are mostly restricted to scalar perturbations. However the equation for axial gravitational perturbations of gravastars is identical to the equation for scalar perturbations in the large $l=m$ limit \cite{paper1}. There are also generic arguments suggesting that the timescale of gravitational perturbations is smaller than the timescale of scalar perturbations for low $m$ \cite{Kokkotas:2002sf}. Thus, scalar perturbations should provide a lower bound on the
strength of the instability.
\begin{widetext}
\begin{center}
\begin{table}[ht]
\centering
\caption{\label{tab:instrotgravaII} WKB results for the instability of rotating gravastars with $r_2=2.2$, $r_1=1.8$ and $M=1$.}
\begin{tabular}{|c||ccccc|}
%
%\hline
\hline \multicolumn{1}{|c}{} & \multicolumn{5}{c|}{ $\tau/M$}\\
\hline
&$J/M^2=0.40$    &$J/M^2=0.60$    &$J/M^2=0.80$    &$J/M^2=0.90$     &$J/M^2=1.0$\\
$m$&$\Omega/\Omega_K=0.33$&$\Omega/\Omega_K=0.49$&$\Omega/\Omega_K=0.65$&$\Omega/\Omega_K=0.74$&$\Omega/\Omega_K=0.82$\\
1&$1.33\t10^{7}$&$2.78\t10^{4}$&$5.99\t10^{3}$&$3.58\t10^{3}$&$2.34\t10^{3}$\\
2&$8.25\t10^{7}$&$1.14\t10^{6}$&$1.11\t10^{5}$&$4.81\t10^{4}$&$2.33\t10^{4}$\\
3&$1.31\t10^{10}$&$5.65\t10^{7}$&$2.25\t10^{6}$&$6.82\t10^{5}$&$2.45\t10^{5}$\\
4&$2.50\t10^{12}$&$2.95\t10^{9}$&$4.81\t10^{7}$&$1.02\t10^{7}$&$2.73\t10^{6}$\\
5&$5.06\t10^{14}$&$1.59\t10^{11}$&$1.02\t10^{9}$&$1.52\t10^{8}$&$3.07\t10^{7}$\\
\hline \hline
\end{tabular}
\end{table}
\end{center}
\end{widetext}
%%%%%%%%%%%%%%%%%%%%%%%%%%%%%%%%%%%%%%%%%%%%%%%%%%%%%%%%%%%%%%%%%%%%%%%%%%%%%%%%%%%%%%%%%%%%%%%%%%%%%%%%%%%%%%%%%%%%%%
\subsubsection{\label{sec:ergoregioninstslowlyrot} Scalar field instability for slowly rotating gravastars: WKB
approach}
%%%%%%%%%%%%%%%%%%%%%%%%%%%%%%%%%%%%%%%%%%%%%%%%%%%%%%%%%%%%%%%%%%%%%%%%%%%%%%%%%%%%%%%%%%%%%%%%%%%%%%%%%%%%%%%%%%%%%%%%
%%%%%%%%%%%%%%%%%%%%%%%%%%%%%%%%%%%%%%%%%%%%%%%%%%%%%%%%%%%%%%%%%%%%%%%%%%%%%%%%%%%%%%%%%%%%%%%%%%%%
%%%%%%%%%%%%%%%%%%%%%%%%%%%%%%%%
Consider now a minimally coupled scalar field in the background of a gravastar. The metric of
gravastars is given by Eq.~(\ref{rotfull}). In the large $l=m$ limit, which is appropriate for
a WKB analysis \cite{cominsschutz, Cardoso:2005gj}, the scalar field can be expanded in spherical armonics, $Y_{l
m}=Y_{l
m}(\theta\,,\phi)$ as
\be \Phi=\sum_{lm}\bar{\chi}_{lm}\exp{\left [-\frac{1}{2}\int
\left (\frac{2}{r}+\frac{f'}{2f}+\frac{B'}{2B}\right )dr
\right]}e^{-i\omega t}Y_{l
m}\,.\label{scalardecomposition} \ee
The functions $\bar{\chi}_{lm}=\bar{\chi}_{lm}(r)$ are determined by the Klein-Gordon
equation which, dropping terms of order ${\cal O}\left (1/m^2 \right)$, yields
\be \bar{\chi}_{lm}''+m^2T(r\,,\Sigma)\bar{\chi}_{lm}=0\,,
\label{equationphigrava} \ee
where $\Sigma\equiv -\omega/m$ and
\beq
T=\frac{B(r)}{f(r)}\left (\Sigma-V_+\right )\left (\Sigma-V_-\right )\,,\hspace{0.5cm}V_{\pm}&=&-\omega \pm \frac{\sqrt{f(r)}}{r}\,.\nonumber
\eeq
Equation (\ref{equationphigrava}) can be shown to be identical for the axial gravitational perturbations of
perfect fluid stars \cite{paper2}. 

The WKB method \cite{cominsschutz} for computing the eigenfrequencies of Eq.~(\ref{equationphigrava}) is in excellent agreement with full numerical results \cite{Cardoso:2005gj}. The quasi-bound unstable modes are determined by
\be
m\int_{r_a}^{r_b}\sqrt{T(r)}dr =\frac{\pi}{2}+n\pi\,,\quad
n=0\,,1\,,2\,,\dots
\ee
and have an instability timescale
\be
\tau=4\exp{\left[2m\int_{r_b}^{r_c}\sqrt{|T|}dr\right]}\int_{r_a}^{r_b}\frac{d}{d\Sigma}
\sqrt{T}dr\,,
\ee
where $r_a$, $r_b$ are solutions of $V_+=\Sigma$ and $r_c$ is determined by the condition
$V_-=\Sigma$.

Table \ref{tab:instrotgravaII} shows the WKB results for the anisotropic pressure model for different values of $J/M^2$. Although the WKB approximation breaks down at low $m$ values, these results
still provide reliable estimates \cite{cominsschutz}. This claim has be verified  with a full numerical integration of the Klein-Gordon equation. The results show that the instability timescale decreases
as the star becomes more compact. Larger values of $J/M^2$ make the star more unstable. The maximum growth time of the instability can be of the order of a few thousand $M$, but it crucially depends on $J$, $\mu$ and $\delta$ \cite{Chirenti:2008pf}. For a large range of parameters this instability is crucial for the star evolution. Gravitational perturbations are expected to be more unstable. Moreover it is worth to notice that the slowly rotating approximation allows only for $\mu<0.5$, while for rotating BHs $0.5<\mu<1$ (see Table~\ref{tab:DATA}). The ergoregion instability being monotonically increasing with $\mu$, we expect that instability timescales for realistic gravastars should be much shorter than the ones computed. For most of the BH mimickers models to be viable we require $J/M^2\sim1$ and $\mu\sim1$. It would be interesting to study whether the ergoregion instability is or is not always effective in this case. Possible future developments include: (i) a full rotating gravastar model, which allows for $\mu>0.5$; (ii) the stability analysis against gravitational perturbations for rotating gravastars; (iii) a gravavastar model which is not strongly dependent on the thickness, $\delta$. 

The ergoregion instability of a rotating boson star is straightforwardly computed following the method described above for spinning gravastars. We refer the reader to \cite{paper1} and we only summarize the results in Table \ref{tab:instrotboson}. The maximum growth time for this boson star model is of the order of $10^6 M$ for ${J}/{GM^2}=0.857658$. Thus the instability seems to be truly effective for rotating boson stars.
\begin{table}[ht]
\centering
\caption{\label{tab:instrotboson} Instability for rotating boson stars with parameters $n=2$,
$b=1.1$, $\lambda=1.0$, $a=2.0$ and different values of $J$ (from \cite{paper1}). The Newton constant is defined as $4\pi G=0.05$.}
\begin{tabular}{|c||ccc|}
%
%\hline
\hline \multicolumn{1}{|c}{} & \multicolumn{3}{c|}{ $\tau/(GM)$}\\
\hline
$m$  & ${J}/{GM^2}=0.5661$&${J}/{GM^2}=0.7307$ &${J}/{GM^2}=0.8577$    \\
1    & $8.847\times 10^2$  &$6.303\times10^3$ &$-$\\
2    & $7.057\times10^3$   &$5.839\times10^4$ &$1.478\times10^6$\\
3    & $6.274\times10^4$   &$9.274.\times10^5$&$2.815\times10^8$\\
4    & $5.824\times10^5$   &$1.603\times10^7$ &$2.815\times10^{10}$\\
5    & $5.554\times10^6$   &$2.915\times10^8$ &$1.717\times10^{12}$\\
\hline \hline
\end{tabular}
\end{table}
%
%%%%%%%%%%%%%%%%%%%%%%%%%%%%%%%%%%%%%%%%%%%%%%%%%%%%%%%%%%%%%%%%%%%%%%%%%%%%%%%%%%%%%%%%%%%%%%%%%%%%%%%%%%%%%%%%%
\section{A toy model for Kerr-like objects}\label{sec2}
%%%%%%%%%%%%%%%%%%%%%%%%%%%%%%%%%%%%%%%%%%%%%%%%%%%%%%%%%%%%%%%%%%%%%%%%%%%%%%%%%%%%%%%%%%%%%%%%%%%%%%%%%%%%%%%%%
This section discusses Kerr-like objects such as particular solutions of rotating wormholes and superspinars. A rigorous analysis of the ergoregion instability for these models is a non-trivial task. Indeed known wormhole
solutions are special non-vacuum solutions of the gravitational field equations, thus their investigation
requires a case-by-case analysis of the stress-energy tensor. Moreover exact solutions of four-dimensional
superspinars are not known. To overcome these difficulties, the following analysis will focus on a simple
model which captures the essential features of most Kerr-like horizonless ultra-compact objects.
Superspinars and rotating wormholes will be modeled by the exterior Kerr metric down to their surface,
where mirror-like boundary conditions are imposed. This problem is very similar to Press and Teukolsky's
``BH bomb'' \cite{bhbombPress, Cardoso:2004nk}, i.e.\ a rotating BH surrounded by a perfectly reflecting
mirror with its horizon replaced by a reflecting surface. For a more detailed discussion see \cite{paper2}.
%%%%%%%%%%%%%%%%%%%%%%%%%%%%%%%%%%%%%%%%%%%%%%%%%%%%%%%%%%%%%%%%%%%%%%%%%%%%%%%%%%%%%%%%%%%%%%%%
\subsubsection{\label{sec:superspinars} Superspinars and Kerr-like wormholes}
%%%%%%%%%%%%%%%%%%%%%%%%%%%%%%%%%%%%%%%%%%%%%%%%%%%%%%%%%%%%%%%%%%%%%%%%%%%%%%%%%%%%%%%%%%%%%%%%
A superspinar of mass $M$ and angular momentum $J=aM$ can be modeled by the Kerr geometry
\cite{Gimon:2007ur}
\begin{widetext}
\be
ds_{\rm Kerr}^2=-\left(1-\frac{2Mr}{\Sigma}\right)dt^2+\frac{\Sigma}{\Delta}dr^2 +
\left[\frac{(r^2+a^2)}{\sin^2\theta} +\frac{2Mr}{\Sigma}a^2 \right]\sin^4\theta d\phi^2-\frac{4Mr}{\Sigma}a\sin^2\theta d\phi dt
+{\Sigma}d\theta^2\,,
\label{kerrmetric}
\ee
\end{widetext}
where $\Sigma=r^2+a^2\cos^2\theta$ and $\Delta=r^2+a^2-2M r$. Unlike Kerr BHs, superspinars have $a>M$ and no horizon. Since the
domain of interest is $-\infty<r<+\infty$, the space-time possesses
naked singularities and closed timelike curves in regions where
$g_{\phi\phi}<0$ \cite{Chandraspecial}. High energy modifications (i.e. stringy corrections) in the vicinity
of the singularity are also expected.

Kerr-like wormholes are described by the metric
\be ds^{2}_{\rm wormhole}=ds^2_{\rm Kerr}+\delta g_{ab}dx^{a}dx^b\,,
\label{worm}
\ee
where $\delta g_{ab}$ is infinitesimal. In general,
Eq.~(\ref{worm}) describes an horizonless object with a excision at
some small distance of order $\epsilon$ from the would-be horizon \cite{Damour:2007ap}. Wormholes require exotic matter and/or divergent
stress tensors, thus some ultra-stiff matter is assumed close to
the would-be horizon. In the following, both superspinars and
wormholes will be modeled by the Kerr metric with a rigid ``wall''
at finite Boyer-Lindquist radius $r_0$, which excludes the
pathological region.\\
%%%%%%%%%%%%%%%%%%%%%%%%%%%%%%%%%%%%%%%%%%%%%%%%%%%%%%%%%%%%%%%%%%%%%%%%%%%%%%%%%%%%%%%%%%%%%%%%%%%%
\subsection{Instability analysis}
%%%%%%%%%%%%%%%%%%%%%%%%%%%%%%%%%%%%%%%%%%%%%%%%%%%%%%%%%%%%%%%%%%%%%%%%%%%%%%%%%%%%%%%%%%%%%%%%%%%%
If the background geometry of superspinars and wormholes is sufficiently close to the Kerr
geometry, its perturbations is determined by the equations of perturbed Kerr BHs \cite{paper2}. Thus the instability of superspinars and wormholes is studied by considering Kerr geometries with
arbitrary rotation parameter $a$ and a ``mirror'' at some Boyer-Lindquist radius $r_0$. Using the
Kinnersley tetrad and Boyer-Lindquist coordinates, it is possible to separate the angular variables
from the radial ones, decoupling all quantities. Small perturbations of a spin-$s$
field are reduced to the radial and angular master equations \cite{teukolsky}
\begin{widetext}
\be
\Delta^{-s}\frac{d}{dr}\left(\Delta^{s+1}\frac{dR_{l
m}}{dr}\right)+ \left[\frac{K^{2}-2is(r-M)K}{\Delta}+4is\omega r -\lambda\right]R_{l m}=0\,,
\label{wave eq separated general}
\ee
\be
\left[(1-x^2){}_sS_{l m,x} \right]_{,x}+
\left[(a\omega x)^2-2a\omega sx+s+{}_{s}A_{l m}-\frac{(m+sx)^2}{1-x^2}\right]{}_sS_{l m}=0\,,
\label{angularwaveeq}
\ee
\end{widetext}
where $x\equiv\cos\theta$, $\Delta=r^2-2Mr+a^2$ and $K=(r^2+a^2)\omega-am$. Scalar, electromagnetic and gravitational perturbations correspond to $s=0$, $\pm1$, $\pm2$ respectively. The separation constants
$\lambda$ and ${}_s A_{l m}$ are related by $\lambda \equiv {}_s A_{l m}+a^2\omega^2-2am\omega$. 
%%%%%%%%%%%%%%%%%%%%%%%%%%%%%%%%%%%%%%%%%%%%%%%%%%%%%%%%%%%%%%%%%%%%%%%%%%%%%%%%%%%%%%%%%%%%%%%
\subsubsection{Analytic results}
%%%%%%%%%%%%%%%%%%%%%%%%%%%%%%%%%%%%%%%%%%%%%%%%%%%%%%%%%%%%%%%%%%%%%%%%%%%%%%%%%%%%%%%%%%%%%%%
%
\begin{figure}[ht]
\begin{tabular}{cc}
\epsfig{file=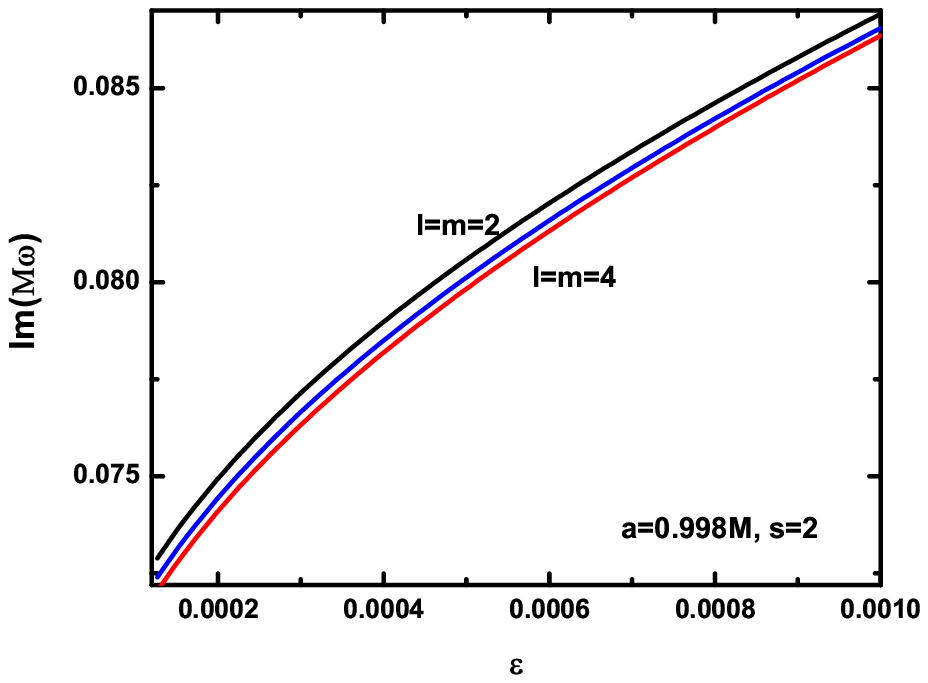,height=140pt,angle=0}\\
\epsfig{file=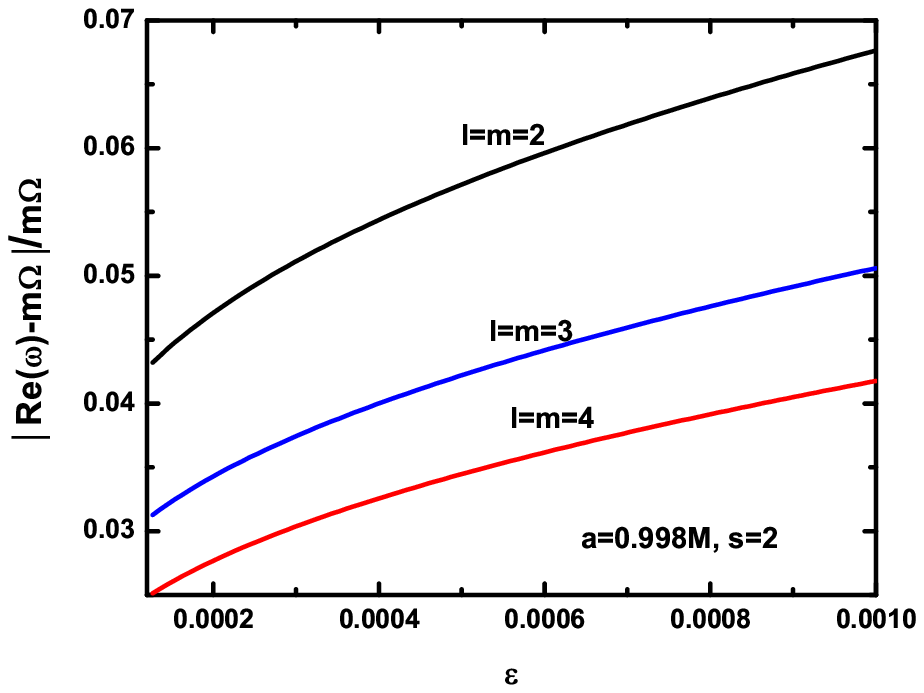,height=140pt,angle=0}
\end{tabular}
\caption{Imaginary and real parts of the characteristic gravitational frequencies for an object with
$a=0.998M$, according to the analytic calculation for rapidly-spinning objects. The mirror location
is at $r_0=(1+\epsilon)r_+$. The real part is approximately constant and close to $m\Omega$, in agreement
with the assumptions used in the analytic approach.}
\label{fig:extremal scalar}
\end{figure}
Following Starobinsky \cite{staro1}, equations (\ref{wave eq separated general})-(\ref{angularwaveeq}) can be analytically solved in the
slowly-rotating and low-frequency regime, $\omega M\ll 1$, and in
the rapidly-spinning regime, where $r_+\sim r_-$ and $\omega \sim m \Omega_h$, where $\Omega_h \equiv a/(2Mr_+)$
is the angular velocity at the horizon. The details of the analytic approximation are described in Ref.\ \cite{paper2}. 
Analytic solutions for a star with $a=0.998M$ are shown in
Fig.~\ref{fig:extremal scalar} where gravitational perturbations are considered. The instability timescale for gravitational perturbations is about five orders
of magnitude smaller than the instability timescale for scalar perturbations.
%%%%%%%%%%%%%%%%%%%%%%%%%%%%%%%%%%%%%%%%%%%%%%%%%%%%%%%%%%%%%%%%%%%%%%%%%%%%%%%%%%%%%%%%%%%%%%%%%%%%
\subsection{Instability analysis: numerical results \label{sec:num}}
%%%%%%%%%%%%%%%%%%%%%%%%%%%%%%%%%%%%%%%%%%%%%%%%%%%%%%%%%%%%%%%%%%%%%%%%%%%%%%%%%%%%%%%%%%%%%%%%%%%%
The oscillation frequencies of the modes can be found from the  canonical form of Eq.~(\ref{wave eq
separated general})
\be
\frac{d^2Y}{dr_*^2}+VY=0\,,
\label{teu canonical}
\ee
where
\beq
Y&=&\Delta^{s/2}(r^2+a^2)^{1/2}R\,,\qquad\nonumber\\
V&=&\frac{K^2-2is(r-M)K+\Delta(4ir\omega
s-\lambda)}{(r^2+a^2)^{2}}-G^2-\frac{dG}{dr_*}\,,\nonumber
\eeq
and $K=(r^2+a^2)\omega-am$, $G=s(r-M)/(r^2+a^2)+r\Delta(r^2+a^2)^{-2}$. The separation constant
$\lambda$ is related to the eigenvalues of the angular equation by $\lambda \equiv {}_s A_{l m}+a^2\omega^2-2am\omega$. The eigenvalues
${}_sA_{lm}$ are expanded in power series of $a\omega$ as \cite{Berti:2005gp}
\be {}_sA_{lm}=\sum_{k=0}f^{(k)}_{slm}(a\omega)^{k}\,.
\label{rel:eigexpans} \ee
Terms up to order $(a\omega)^2$ are included in the calculation.  Absence of ingoing waves at infinity
implies
\be
Y\sim r^{-s}e^{i\omega r_*}\,.
\label{asymp sol}
\ee
Numerical results are obtained by integrating Eq.~(\ref{teu canonical}) inward from a large distance
$r_{\infty}$. The integration is performed with the Runge-Kutta method with fixed $\omega$ starting
at $M r_\infty = 400$, where the asymptotic behavior (\ref{asymp sol}) is imposed. (Choosing a
different initial point does not affect the final results.) The numerical integration is stopped at
the radius of the mirror $r_0$, where the value of the field $Y(\omega,r_0)$ is extracted. The
integration is repeated for different values of $\omega$ until $Y(\omega,r_0)=0$ is obtained with the
desired precision. If $Y(\omega,r_0)$ vanishes, the field satisfies the boundary condition for
perfect reflection and $\omega=\omega_0$ is the oscillation frequency of the mode.
%%%%%%%%%%%%%%%%%%%%%%%%%%%%%%%%%%%%%%%%%%%%%%%%%%%%%%%%%%%%%%%%%%%%%%%%%%%%%%%%%%%%%%%%%%%%%%%
\subsubsection{Objects with $a<M$}
%%%%%%%%%%%%%%%%%%%%%%%%%%%%%%%%%%%%%%%%%%%%%%%%%%%%%%%%%%%%%%%%%%%%%%%%%%%%%%%%%%%%%%%%%%%%%%%
%%%%%%%%%%%
\begin{center}
\begin{table}[ht]
\centering
\caption{\label{tab:inststarbomb} Characteristic frequencies and
instability timescales for a Kerr-like object with $a=0.998M$. The
mirror is located at $\epsilon=0.1$, corresponding to the compactness $\mu\sim0.9\mu_{\text{Kerr}}$.}
\begin{tabular}{|c||cc|}
%
%\hline
\hline \multicolumn{1}{|c}{} & \multicolumn{2}{c|}{ $({\rm Re}(\omega)M\,,{\rm Im}(\omega)M)$}\\
\hline
$l=m$  & $s=0$                    &$s=2$\\
1    & $(0.1120\,,0.6244\times 10^{-5})$          &$-$    \\
2    & $(0.4440\,,0.5373\times 10^{-5})$         &$ (0.4342\,,0.2900)$    \\
3    & $(0.7902\,,0.1928\times 10^{-5})$        &$(0.7803\,,0.2977)$    \\
4    & $(1.1436\,,0.5927\times 10^{-6})$        &$ (1.1336\,,0.3035)$    \\
\hline \hline
\end{tabular}
\end{table}
\end{center}
The regime $a<M$ requires a surface or mirror at $r_0=r_+(1+\epsilon)>r_+$. Thus the compactness is $M/r_0\sim (1-\epsilon)M/r_+$ and, in the limit $\epsilon\rightarrow0$, it is infinitesimally close to the compactness of a Kerr BH. Numerical results for scalar and gravitational
perturbations of objects with $a<M$ are summarized in Table \ref{tab:inststarbomb} and are in agreement with the analytic results \cite{paper2}. The instability is weaker for larger $m$. This result holds also
for $l\ne m$ and $s=0$, $\pm1$ and $\pm2$. The minimum instability timescale is of order $\tau \sim 10^5M$ for a wide range of mirror locations. Figure \ref{fig:severalrotation} shows the results for gravitational perturbations. Instability timescales are of the order of $\tau \sim 2\div6M$. Thus gravitational perturbations lead to an instability about five
orders of magnitude stronger than the instability due to scalar perturbations (see Table
\ref{tab:inststarbomb}). Figure \ref{fig:severalrotation} shows that the ergoregion instability remains
relevant even for values of the angular momentum as low as $a=0.6M$.
\begin{widetext}
\begin{center}
\begin{figure}[ht]
\begin{tabular}{cc}
\epsfig{file=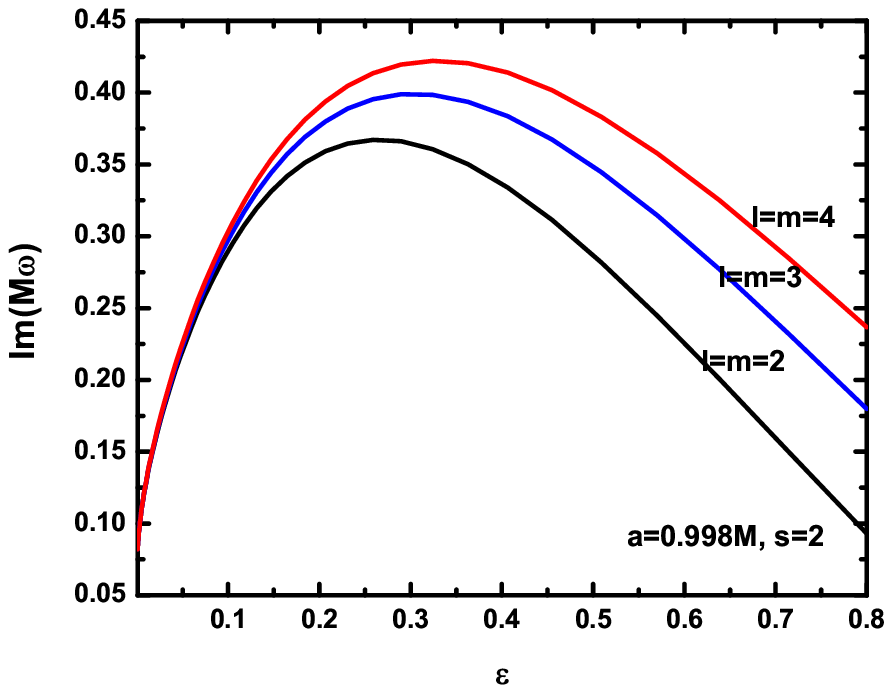,height=140pt,angle=0} &
\epsfig{file=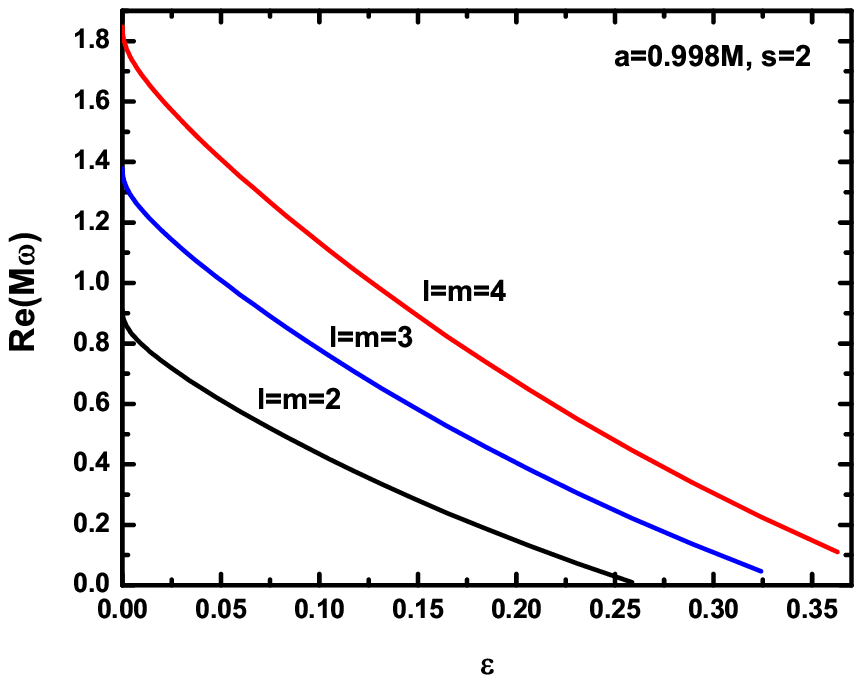,height=140pt,angle=0}\\
\epsfig{file=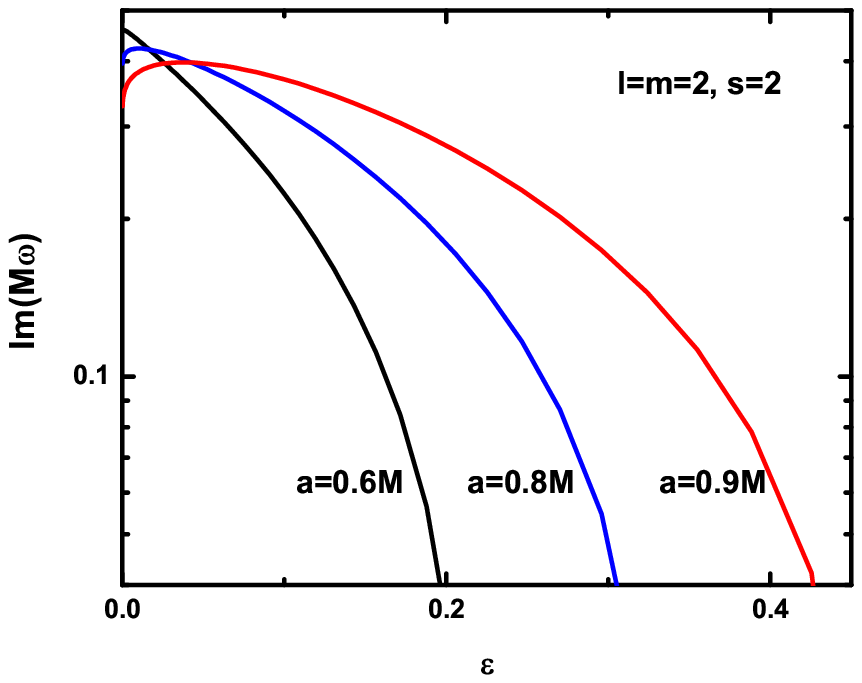,height=140pt,angle=0}&
\epsfig{file=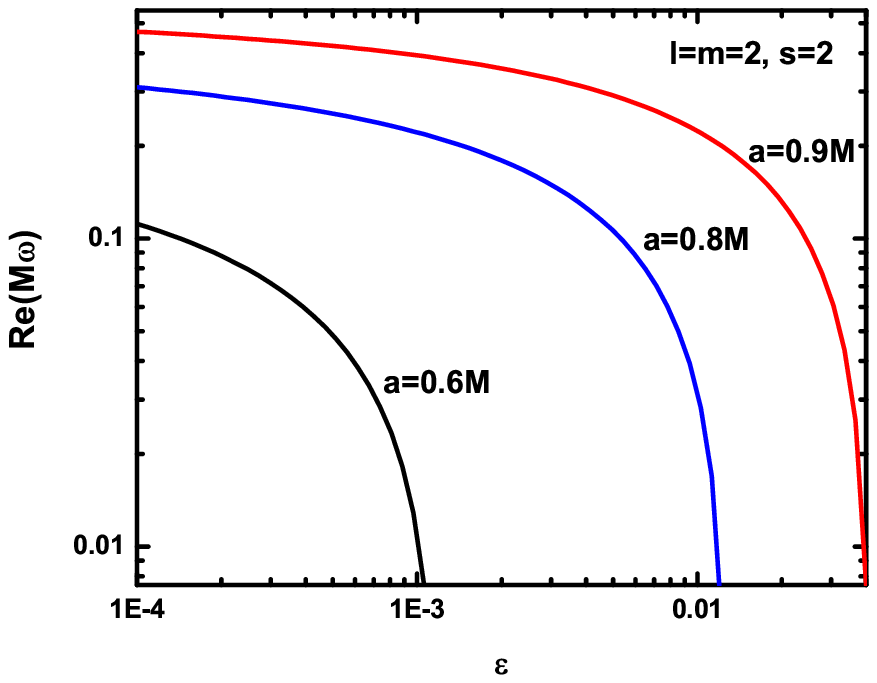,height=140pt,angle=0}
\end{tabular}
\caption{Details of the instability for gravitational perturbations, for different $l=m$ modes and $a/M=0.998$
(top panels) and for $l=m=2$ and different $a/M<1$.}
\label{fig:severalrotation}
\end{figure}
\end{center}
\end{widetext}
%%%%%%%%%%%%%%%%%%%%%%%%%%%%%%%%%%%%%%%%%%%%%%%%%%%%%%%%%%%%%%%%%%%%%%%%%%%%%%%%%%%%%%%%%%%%%%%
\subsubsection{Objects with $a>M$}
\begin{figure}[ht]
\begin{tabular}{cc}
\epsfig{file=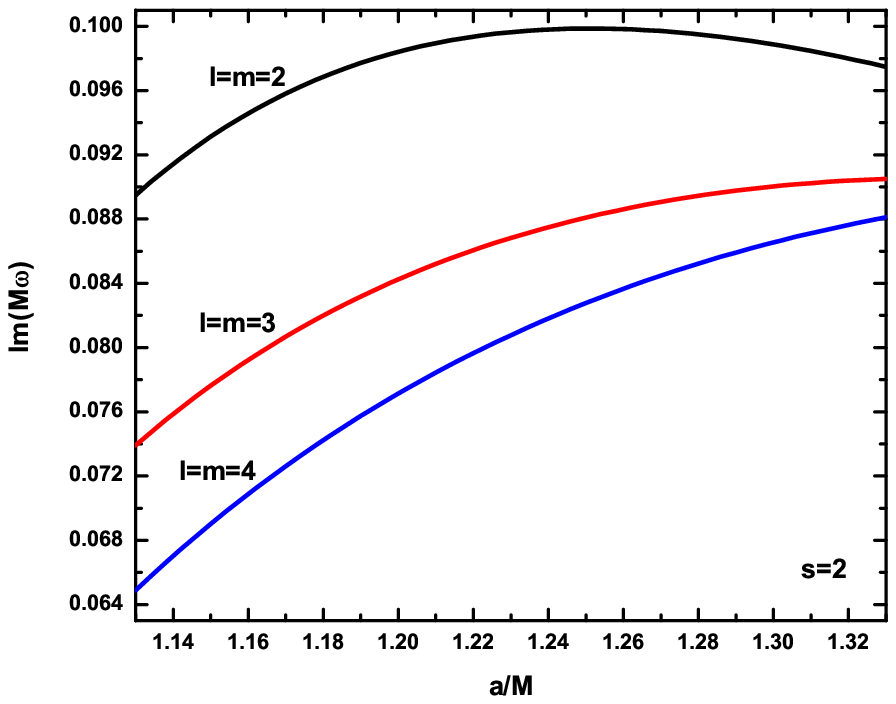,height=.37\textwidth,angle=0}\\
\epsfig{file=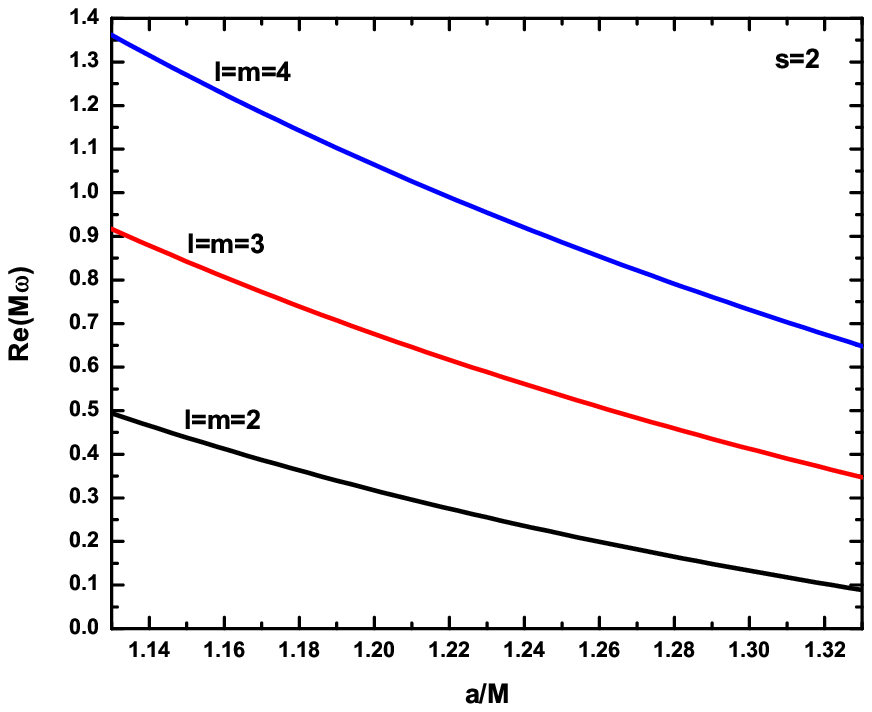,height=.35\textwidth,angle=0}
\end{tabular}
\caption{The fundamental $l=m=2,3,4$ modes of an object spinning above the Kerr bound as function of
rotation. The surface is located at $r_0/M=0.001$.}
\label{fig:super}
\end{figure}

%%%%%%%%%%%%%%%%%%%%%%%%%%%%%%%%%%%%%%%%%%%%%%%%%%%%%%%%%%%%%%%%%%%%%%%%%%%%%%%%%%%%%%%%%%%%%%%
Objects with $a>M$ could potentially describe superspinars.
Several arguments suggest that objects rotating above the Kerr
bound are unstable. Firstly, extremal Kerr BHs are marginally
stable. Thus the addition of extra rotation should lead to
instability. Secondly, fast-spinning objects usually take a
pancake-like form \cite{Emparan:2003sy} and are subject to the
Gregory-Laflamme instability \cite{Gregory:1993vy,Cardoso:2006ks}.
Finally, Kerr-like geometries, like naked singularities, seem to
be unstable against a certain class of gravitational perturbations
\cite{Cardoso:2006bv, Dotti:2006gc} called
algebraically special perturbations \cite{Chandraspecial}. For objects with $a>M$ the surface
or mirror can be placed anywhere outside $r=0$. In general the
instability is as strong as in the $a<M$ regime. An example in
shown in Fig.\ \ref{fig:super} for the surface at $r_0/M=0.001$.
This result confirms other investigations suggesting that
ultra-compact objects rotating above the Kerr bound are unstable
\cite{Dotti:2008yr}.
\section{Conclusion\label{sec:discussion}}
%%%%%%%%%%%%%%%%%%%%%%%%%%%%%%%%%%%%%%%%%%%%%%%%%%%%%%%%%%%%%%%%%%%%%%%%%%%%%%%
We investigated the ergoregion instability of some ultra-compact, horizonless objects which can mimick the spacetime of a rotating black hole. We studied some of the most viable BH mimickers: gravastars, boson stars, wormhole and superspinars.

If rotating, boson stars and gravastars may develop ergoregion instabilities. Analytical and numerical
results indicate that these objects are unstable against scalar field perturbations for a large range of the parameters. Slowly rotating gravastars can develop an ergoregion depending on their angular momentum, their compactness and the thickness of their intermediate region. In a recent work \cite{Chirenti:2008pf} it has pointed out that slowly rotating gravastars may not develop an ergoregion. In the formation of the ergoregion for rotating gravastars an important role is played by the thickness (see Figure \ref{fig:delta}) which is not easily detectable. Thus further investigations are needed to better understand the ergoregion formation in physical resonable gravastar models.

The instability timescale for both boson stars and gravastars can be many orders of magnitude stronger than the instability timescale for ordinary stars with uniform density. In the large $l=m$ approximation,
suitable for a WKB treatment, gravitational and scalar perturbations have similar instability
timescales. In the low-$m$ regime gravitational perturbations are expected to have even shorter
instability timescales than scalar perturbations. Instability timescales can be as low as $\sim0.1$ seconds for a $M=1 M_\odot$ objects and about a week for supermassive BHs, $M=10^6 M_{\odot}$, monotonically decreasing for larger rotations and a larger compactness.

The essential features of wormholes and superspinars have been captured by a simple
model whose physical properties are largely independent from the dynamical details of the gravitational
system. Numerical and analytic
results show that the ergoregion instability of these objects is extremely strong for any value of their
angular momentum, with timescales of order $10^{-5}$ seconds for a $1 M_\odot$ object and $10$ seconds for a
$M=10^6 M_{\odot}$ object. Therefore, high rotation is an indirect evidence for horizons. 

Although further studies are needed, the above investigation suggests that exotic objects without event horizon are likely to be ruled out as viable candidates for astrophysical ultra-compact objects. This strengthens the role of BHs as candidates for astrophysical observations of rapidly spinning compact objects.
%%%%%%%%%%%%%%%%%%%%%%%%%%%%%%%%%%%%%%%%%%%%%%%%%%%%%%%%%%%%%%%%%%%%%%%%%%%%%%%%%%%%%%%%%%%%%%%%%%%%%%%%%%%%%%%%%
%%%%%%%%%%%%%%%%%%%%%%%%%%%%%%%%%%%%%%%%%%%%%%%%%%%%%%%%%%%%%%%%%%%%%%%%%%%%%%%%%%%%%%%%%%%%%%%%%%%%%%%%%%%%%%%%%
%%%%%%%%%%%%%%%%%%%%%%%%%%%%%%%%%%%%%%%%%%%%%%%%%%%%%%%%%%%%%%%%%%%%%%%%%%%%%%%
\section*{Acknowledgements}
% %%%%%%%%%%%%%%%%%%%%%%%%%%%%%%%%%%%%%%%%%%%%%%%%%%%%%%%%%%%%%%%%%%%%%%%%%%%%%%%
%The work presented above has been done in collaboration with Vitor Cardso, Mariano Cadoni and Marco Cavagli\`a.
The authors warmly thank Matteo Losito for interesting discussions and for sharing some of his results. This work is supported by Funda\c{c}\~{a}o para a Ci\^{e}ncia e Tecnologia (FCT) - Portugal through project PTDC/FIS/64175/2006 and
by the National Science Foundation through LIGO Research Support grant NSF PHY-0757937. 
%%%%%%%%%%%%%%%%%%%%%%%%%%%%%%%%%%%%%%%%%%%%%%%%%%%%%%%%%%%%%%%%%%%%%
%%%%%%%%%%%%%%%%%%%%%%%%%%%%%%%%%%%%%%%%%%%%%%%%%%%%%%%%%%%%%%%%%%%%%%%%%%%%%%%%%%%%%%%%%%%%%%%%%%%%%%%%%%%%%%%%%

\end{document}